\documentclass[12pt]{article}

\renewcommand{\title}[1]{
\begin{center} \Large \bf #1 \end{center}
}

\renewcommand{\author}[3]{
 \begin{center} #1 \\
  #2 \\
  {\small E-mail: \texttt{#3}}
 \end{center}
\addvspace{\baselineskip}
}

\usepackage{amssymb}
\usepackage{amsmath}

\usepackage{amsthm}
\newtheorem{thm}{Theorem}[section]

\newtheorem{lem}[thm]{Lemma}

\theoremstyle{definition}

\theoremstyle{remark}

\font\mybb=msbm10 at 12pt
\def\bb#1{\hbox{\mybb#1}}

\font\mybb=msbm10 at 12pt

\makeatletter
\@addtoreset{equation}{section}

\makeatother
\begin{document}

\baselineskip 5mm

\title{Are Vortex Numbers Preserved?}

\author{Yoshiaki Maeda, Akifumi Sako${}^\dagger$}{Department of Mathematics,
Faculty of Science and
 Technology, Keio University\\
3-14-1 Hiyoshi, Kohoku-ku, Yokohama 223-8522, Japan }
{maeda@math.sci.keio.ac.jp, ${}^\dagger$sako@math.sci.keio.ac.jp }

\vspace{1cm}

\abstract{
We study noncommutative vortex solutions that 
minimize the action functional of
the Abelian Higgs model in 2-dimensional 
noncommutative Euclidean space.
We first consider vortex solutions which are deformed from solutions 
defined on commutative Euclidean space to
the noncommutative one.
We construct solutions whose vortex numbers are unchanged
under the noncommutative deformation.
Another class of noncommutative vortex solutions via a Fock
space representation is also studied.
}
% main text
%%%%%%%%%%%%%%%%%%%%%%%%%%%%%%%%%%%%%%%%%%%%%%%%%%%%%%%%%%%%%%%%
\section{Introduction}
\label{section1}
%%%%%%%%%%%%%%%%%%%%%%%%%%%%%%%%%%%%%%%%%%%%%%%%%%%%%%%%%%%%%%%%%%%%%%%%%%%

\newcommand{\bra}[2]{\left<#1,#2\right|}
%%%%%%%%%%%%%%%%%%%%%%%%%%%%%%%%%%%%%%%%%%%%%%%%%%%%%%%%%%%%%%%%%%%%%%%%%%%
%%%%%%%%%%%%%%%%%%%%%%%%%%%%%%%%%%%%%%%%%%%%%%%%%%%%%%%%%%%%%%%%%%%%%%%%%%%

In the noncommutative Euclidean space, 
the instanton number is given by an integer which does not depend on the noncommutative
parameter, for the instanton
solutions given by ADHM construction
\cite{sako2,sako3,Furuuchi1,Furuuchi2,Tian}.
Because of these observations, one can ask
``Are topological charges 
unchanged when we deform the space from Euclidean space
to noncommutative Euclidean space?".
To answer this question,
we investigate a two dimensional Abelian
Higgs model.
Solutions of the Bogomol'nyi equations in this model 
are called vortex solutions, and the vortex solutions 
minimize the action functional of the
Abelian Higgs model.\\

In this paper, we study vortex solutions in noncommutative Euclidean space.
We consider solutions which are deformations of vortex solutions defined on 
commutative Euclidean space and ask if the vortex number changes under the noncommutative deformation.
In this paper, we use Taubes' solution \cite{taubes}
as the vortex solution before undergoing deformation. 
The main purpose of this paper is to show that vortex numbers of
vortex solutions are unchanged under this noncommutative deformation.

The organization of this article is as follows.
In the next section, we review some results about the two 
dimensional Abelian Higgs model and vortices, and
we lay out the notation of this article.
In section \ref{NCAHM},
we define and discuss the 
noncommutative deformation of the Abelian Higgs model.
%Some facts of the deformed theory are discussed, too.
In section \ref{section vortex}, 
we investigate the noncommutative vortex solutions deformed from the
commutative vortex solutions and their vortex numbers.
Our main claim is that the vortex number is unchanged.
At first, we show that the vortex number is unchanged
under certain conditions.
Next, we solve the noncommutative vortex equations, and we show that the solutions 
satisfy these conditions.
In section \ref{anothe type}, another type of solution is
treated.
These solutions are not given by deformations of commutative vortex
solutions, but are constructed using the Fock space representation.
We show that one of the solutions is given by
a bounded function.

%%%%%%%%%%%%%%%%%%%%%%%%
\section{ Taubes' Vortex Solutions} \label{sect2}
%%%%%%%%%%%%%%%%%%%%%%%

We summarize the $U(1)$ gauge theory
in commutative $\mathbb{R}^{2}$.
The gauge theory is defined by an action functional invariant under 
the gauge transformation. 
For example, the gauge symmetry is defined by the Higgs field.
Higgs field ${\phi}$, a complex scalar field. %of $SO(2)$
Let $G$ be the group of gauge transformations associated to $U(1)$.
For $ g \in G$, the gauge transformation is defined as 
$$ \phi \rightarrow {g}{\phi} .$$
Noting that $\partial_{\mu} {\phi}$ is not covariant 
under this gauge transformation,
%Instead of $\partial_{\mu} {\phi}$,
%let us define 
we introduce the covariant derivative operator by
\begin{equation}
 \nabla_{\mu}:=\partial_{\mu}-i A_{\mu}\;,\label{EQ*:covariant-derivative}
\end{equation}
where $A_{\mu}$ are the components of a local 1-form (a section of the cotangent bundle on ${\bb R}^2$).
Its gauge transformation is defined by 
\begin{equation}
{\textsf{A}} \rightarrow  i g \textsf{d}  g^{-1} + {\textsf{A}} \ .
\end{equation}
Here ${\textsf{A}} :=A_{\mu} dx^{\mu} \in \Omega^1$.
Under the gauge transformation,
\begin{equation}
 {\nabla}_{\mu} {\phi}
  ={\partial}_{\mu} {\phi}- i{A}_{\mu}  {\phi}
\end{equation}
is covariant.

%%%%%%%%%%%%%%%%%%%%%%%%%%%%%%%%%%%%%%%%%%%%%%%%%%%%%%%%%%%%%%%%%%%%%%%%%%%

%%%%%%%%%%%%%%%%%%%%%%%%%%%%%%%%%%%%%%%%%%%%%%%%%%%%%%%%%%%%%%%%%%%%%%%%%%%
%

For later convenience,
we introduce complex coordinates for ${\mathbb R}^2$ and $A_{\mu}$.
%Here, we list the expression used in this article.
On ${\mathbb R}^2$, we use the following complex coordinates ;
\begin{equation}
  z  =\frac{1}{\sqrt{2}}(x^1+ix^2)\ , \ 
  \bar{z}=\frac{1}{\sqrt{2}}(x^1-ix^2)\ ,
% \begin{cases}
%  x^1=\frac{1}{\sqrt{2}}(z+\bar{z})\\
%  x^2=\frac{1}{\sqrt{2}i}(z-\bar{z})\\
% \end{cases}.
\end{equation}
and define differential operators $\partial , \bar{\partial}$ by
\begin{equation}
% \begin{cases}
  \partial=\frac{1}{\sqrt{2}}(\partial_1-i\partial_2)\ , \ 
  \bar{\partial}=\frac{1}{\sqrt{2}}(\partial_1+i\partial_2) \ ,
% \end{cases}
% ,\hspace{1cm}
% \begin{cases}
%  \partial_1=\frac{1}{\sqrt{2}}(\partial+\bar{\partial})\\
%  \partial_2=\frac{i}{\sqrt{2}}(\partial-\bar{\partial})
% \end{cases}.
\end{equation}
and define complex gauge fields by
\begin{equation}
% \begin{cases}
  A
  =\frac{1}{\sqrt{2}}(A_1-iA_2)\ , \ 
  \bar{A}
  =\frac{1}{\sqrt{2}}(A_1+iA_2) \ .
% \end{cases}
% ,\hspace{1cm}
% \begin{cases}
%  A_1=\frac{1}{\sqrt{2}}(A+\bar{A})\\
%  A_2=\frac{i}{\sqrt{2}}(A-\bar{A})
% \end{cases}.
\end{equation}
The gauge transformations are 
\begin{equation}
A \rightarrow i g  \partial g^{-1} +   A  
\ , \ \ \ 
\bar{A} \rightarrow -i \bar{\partial}g   g^{-1} + 
 \bar{A} \ .
\end{equation}
The curvature for the connection ${\textsf A}$ is
expressed in the coordinates $z, \bar{z}$ as  
\begin{equation}
 F_{z z}=F_{\bar{z} \bar{z}}=0\ , \ 
 F_{z \bar{z}}=iF_{12}=\partial \bar{A}-\bar{\partial} A
\ .\nonumber
\end{equation}
We define the magnetic field $B$ by
\footnote{
We can treat our solutions as the soliton solutions
in the 2+1 dimensional theory.
%the solution that minimize the action functional of 
The static energy density of the 
gauge field is described by
the magnetic field $B$.}
$$B := -i F_{z \bar{z}} \ .$$

Using this representation, the covariant derivatives of the Higgs fields are
\begin{eqnarray}
D \phi = ( \partial -i A ) \phi  \ &,& \ \ \
\bar{D} \phi = (\bar{\partial} - i \bar{A} )  \phi \ , \\
D \bar{\phi}  = \partial \bar{\phi} + i \bar{\phi} A  \ &,& \ \ \
\bar{D} \bar{\phi} = \bar{\partial}\bar{\phi} + i \bar{\phi} \bar{A}
\ \ . \label{Dphi}
\end{eqnarray}
It is worth commenting on the order of the fields.
In the commutative case, the order is irrelevant
e.g. 
$\bar{\phi} A =A \bar{\phi}$ , and so on.
But $\bar{\phi} A \neq A \bar{\phi}$ in the noncommutative case.
Therefore, we use above expression in (\ref{Dphi}).

%%%%%%%%%%%%%%%%%%%%%%%%%%%%%%%%%%%%%%%%%%%%%%%%%%%%%%%%%%%%%%%%%%%%%
The functional studied in this paper
(the static energy functional for the 2+1 dimensional Abelian
Higgs model \cite{Ginzburg_Landau}) is given by
\begin{eqnarray}
S= \int d^2 z \ \Big\{ -\frac{1}{2}  (F_{z \bar{z}})^2 
+ D  \phi  \bar{D}  \bar{\phi} + \bar{D}  \phi  D  \bar{\phi}
+ \frac{1}{2} (\phi \bar{\phi} -1 )^2 \ \Big\} \ .
\end{eqnarray}
Here $d^2 z= d^2 x$.
We can regard this functional as the action functional
of 2 dimensional Abelian Higgs model
\footnote{In the following, we do not distinguish the energy functional
in 2+1 dimensional theory from the 2 dimensional action functional.
For example, a static solution that minimizes the 2+1 dimensional 
energy functional is identified with a solution that
minimizes the 2 dimensional action functional.}.
$S$ can be rewritten as
\begin{eqnarray}
S = S_T 
+ \int d^2 z \ \left\{ 2 \bar{D}  \phi  D  \bar{\phi} 
+ \frac{1}{2} ( B+ (\phi  \bar{\phi} -1 ))^2 \right\}
\ \ , \\
S_T := \int  \ \Big[ \frac{1}{2} \left\{
\textsf{d} ( i \phi  \textsf{d}_{\textsf{A}}  \bar{\phi} - i (\textsf{d}_{\textsf{A}}
  \phi ) \bar{\phi} )
\right\} + \textsf{B} \ \Big] \ \ .
\end{eqnarray}
Here, $\textsf{d}_{\textsf{A}}= \textsf{d} -i\textsf{A}$ and 
$\textsf{B}= B dx^1 \wedge dx^2$ .
$S_T$ is a topological term.
Therefore the vortex equations are given by
\begin{equation}
\bar{D}  \phi = ( \bar{\partial} - i \bar{A} ) \phi = 0 
\ \ , \ \ 
B+ \phi  \bar{\phi} -1 = 0 \ \ . \label{BPS_1}
\end{equation}
Solutions of these Bogomol'nyi 
equations (\ref{BPS_1}) minimize the energy functional.
%%%%%%% (rev) remove : without the topological term.
We call these equations vortex equations and their solutions 
are called vortex solutions.
We list some facts concerning vortex solutions.

\begin{thm}[Taubes, \cite{taubes}] \label{com_vortex_number}
Let $(A_0 , \phi_0 )$ be a smooth solution of (\ref{BPS_1}). 
The vortex number, 
\begin{equation}
N_0 := \frac{1}{2 \pi} \int d^2x B_0 \ , \label{vortex number}
\end{equation}
is an integer equal to
the winding number of $\lim_{|z| \rightarrow \infty} \phi_0 $,
where $B_0 := B(A_0)$.
Therefore, if $N_0 \neq 0$ then $\lim_{|z| \rightarrow \infty} \phi_0 $
must have a zero and 
$arg\:  \phi_0$ cannot be smooth.
\end{thm}
We will focus on noncommutative deformations of this theorem
in section \ref{section vortex}.

{}To describe local expressions for the Higgs field near
the zero points, let us introduce some symbols.
Let $(A_0 , \phi_0 )$ be a smooth solution of (\ref{BPS_1}).
Define the zero set $Z(\phi_0)$ by
\begin{equation}
Z(\phi_0)= \{ z \in \mathbb{C} | \phi_0 (z) =0 \}
\end{equation}

\begin{thm} [Taubes, \cite{taubes}]
\label{Taubes-1} %\\
%{\sl
Let $(A_0 , \phi_0 )$ be a smooth, locally $L^2$ solution of (\ref{BPS_1})
of vortex number $N$.
%and locally square integrable.
Then %$(A_0 , \phi_0 )$ is  
there exist $N$ points $\{ z_1 , \dots , z_N \} $  in 
$\mathbb{C} $, such that
\begin{equation}
Z(\phi_0)=\{ z_1 , \dots , z_N \} .
\end{equation}
There is a neighborhood of each $z_a$ in which
\begin{equation}
\phi_0 (z) = (z- z_a)^{n_a} h_a (z) \ ,
\end{equation}
where $n_a$ is the multiplicity of the 
point $z_a$ in $\{ z_1 , \dots , z_N \}$, and
$ h_a (z)$ is a $C^\infty $, nonvanishing function.
%}
\end{thm}

Finally, we list the following useful formula.
\begin{thm}[Taubes, \cite{taubes}]
Let $(A_0 , \phi_0 )$ be a smooth, finite action solution to the
equations (\ref{BPS_1}). Then for any $\epsilon > 0$, 
there exists $M(\epsilon ) < \infty $ such that
\begin{equation}
0 < \frac{1}{2}(1-|\phi_0(x)|^2)< M(\epsilon ) e^{-r(1-{\epsilon})}
\ , \label{taubes2}
\end{equation}
where $r=|x|$.
\end{thm}
{}From (\ref{taubes2}), the asymptotic behaviors 
of the $(A_0 , \phi_0 )$ for large radius $r$ are given by
\begin{eqnarray}
|\phi_0 | &\sim& 1-C e^{-r(1-{\epsilon})}  \label{asym_phi0}\\
|\partial \phi_0 | &\sim& |\bar{\partial} \phi_0 | \sim C' \frac{1}{r} \nonumber \\
| A_0 | &\sim& {C''} \frac{1}{r} \ .  \label{asym_A0}
\end{eqnarray}
Here, $C , C' , {C''}$ are some constants.

In the following, we investigate the noncommutative deformations
of this theory.
In particular, we will carefully discuss whether
the vortex number is constant.

%%%%%%%%%%%%%%%%%%%%%%%%%%%%%%%%%%%%%%%%%%%%%%%%%%%%%%%%%%%%%%%%%
%%%%%%%%%%%%%%%%%%% 1 %%%%%%%%%%%%%%%%%%%%%%%%%%%%%%%%%%%%%%%%%%%
%%%%%%%%%%%%%%%%%%%%%%%%%%%%%%%%%%%%%%%%%%%%%%%%%%%%%%%%%%%%%%%%%
\section{The Noncommutative Abelian Higgs Model}\label{NCAHM}

%%%%%%%%%%%%%%%%%%%%%%%%%%%%%%%%%%%%%%%%%%%%%%%%%%%%%%%%%%%%%%%%%%%%%%%%%%%

%\newcommand{\bra}[2]{\left<#1,#2\right|}
%%%%%%%%%%%%%%%%%%%%%%%%%%%%%%%%%%%%%%%%%%%%%%%%%%%%%%%%%%%%%%%%%%%%%%%%%%%
%%%%%%%%%%%%%%%%%%%%%%%%%%%%%%%%%%%%%%%%%%%%%%%%%%%%%%%%%%%%%%%%%%%%%%%%%%%

In this section, we deform the Abelian 
Higgs model introduced in the previous section
via the Moyal product \cite{Moyal}.
The vortex equations 
and their solutions are also
deformed.

%%%%%%%%%%%%%%%%%%%%%%%%%%%%%%%%%%%%%%%%%%%%%%%%%%%%%%%%%%%%%%%%%%%
\subsection{The Noncommutative U(1) Gauge Transformation} \label{ncU(1)}

At first, let coordinates of noncommutative Euclidean space
$\mathbb{R}^{2}_{\theta} $ be $x^{\mu}\ , \ \mu =1,2 $ ,
with commutation relations 
\begin{equation}
 [x^{\mu},x^{\nu}]=i\theta \epsilon^{\mu\nu},
 \mu , \nu = 1,2 \ \ ,
 \label{EQ:x-com-rel}
\end{equation}
where 
$\epsilon^{\mu\nu}=-\epsilon^{\nu\mu} \ , \ (\epsilon^{12}=1)$ is an anti-symmetric tensor
and $\theta$ is a parameter called the
noncommutative parameter.
There are several representations of $\mathbb{R}^{2}_{\theta} $.
In this section, we use the Moyal product \cite{Moyal}.
The Moyal product is defined as an integral form
\begin{eqnarray}
 f(x)* g(x)
  &:=& \frac{1}{2\pi\theta} \int_{\mathbb{R}^{2}} \int_{\mathbb{R}^{2}} 
f(y) g(z)
e^{2iS(x,y,z)/\theta} dy dz\ ,
\label{Moyal add}
\\
\mbox{where }\ S(x, y, z)&=&(x,Jy)+(y,Jz)+(z, Jx) \ \mbox{and} 
\ J=
\left( 
\begin{array}{cc}
0& 1 \\
-1&0
\end{array} \right) \nonumber
\end{eqnarray}
for a suitable class of functions on $\mathbb{R}^{2}$
(e.g. subclass of Schwarz functions).  
For our
purpose of this paper to find asymptotic solutions of 
deformed vortex solutions, we
consider the formal version of (\ref{Moyal add}) as follows:
%The Moyal product (star product) is defined by
\begin{eqnarray}
 f(x)* g(x)
  &:=&f(x)\exp\left(\frac{i}{2}\overleftarrow{\partial}_{\mu}
	      \theta \epsilon^{\mu\nu}\overrightarrow{\partial}_{\nu}\right)g(x)
  \nonumber\\
 &=&f(x)g(x)+\sum_{n=1}^{\infty}\frac{1}{n!}f(x)
  \left(\frac{i}{2}\overleftarrow{\partial}_{\mu}
 \theta \epsilon^{\mu\nu}\overrightarrow{\partial}_{\nu}\right)^ng(x)\;.
 \nonumber
\end{eqnarray}
Here $\overleftarrow{\partial}_{\mu}$ is a derivative 
operator for $f(x)$ and 
$\overrightarrow{\partial}_{\nu}$ is for $g(x)$.
Though we are working on  formal vortex solutions 
in section\ref{section vortex}, it 
will be an interesting problem to consider nonformal vortex solutions.

Let us summarize the $U(1)$ gauge theory
on $\mathbb{R}^{2}_{\theta} $.
%Yang-Mills theory is defined by the gauge 
%symmetry. Let us introduce Higgs field to define gauge symmetry.
As in section \ref{sect2}, that is 
the Higgs field is ${\phi}$ %is scalar field. %of $SO(2)$
and the gauge transformation group is $G$ .
%be the gauge transformation group associated with $U(N)$.
For $ g \in G$, gauge transformations are defined as 
$$ \phi \rightarrow {g}*{\phi} .$$
%%%%%%%%%%%%%%%%%%%%%%%%%%%%%%%%%%%%%%%%%%%%%%%%%%%%%%%%%%%%%%
%%%%%%%%%%%%%%%%%%%%%%%%%%%%%%%%%%%%%%%%%%%%%%%%%%%%%%%%%%%%%%%%%%%%%%%%%%
We should comment here that the noncommutative $U(1)$ gauge symmetry is itself deformed from the commutative case. Let $U(x , \theta) \in G$ and $\bar{U}$ 
be the complex conjugate of $U$, 
where $G$ is the gauge transformation 
group of U(1), such that
\begin{equation}
U * \bar{U} = \bar{U}* U =1 \label{U(1)}
\end{equation}
We can expand $U$ as $U(x , \theta)= \sum_{k=0} U_k (x) \theta^k$. 
Then the unitary equation (\ref{U(1)}) is
equivalent to 
\begin{eqnarray}
U_0 \bar{U}_0 &=& 1 \nonumber \\
U_0 \bar{U}_1 +  U_1 \bar{U}_0 +
\frac{1}{2}(\partial U_0 \bar{\partial} \bar{U}_0 -
 \bar{\partial}U_0 \partial \bar{U}_0 ) &=& 0 \nonumber \\
{} &\vdots& {} \nonumber \\
\sum_{0\le l\le m\le p\le k} \!\!\!\!\!
\partial^{m-l}\bar{\partial}^l U_{p-m} 
\bar{\partial}^{m-l} \partial^l \bar{U}_{k-p} 
\frac{(-1)^l \theta^k }{l!(m-l)! 2^m}&=&0 \nonumber \\
{} &\vdots& {} \nonumber 
\end{eqnarray}
One degree of freedom of $U_k$ is determined by solving the above unitary 
equation, and
then only one degree for each $U_k$ is left for the gauge transformation
parameter.
When the expansion of $\phi$ is given by $\sum \phi_k \theta^k$,
the gauge transformation for each $\phi_k$ is
\begin{eqnarray}
\phi & \rightarrow & \phi' = \sum_k {\phi'}_k \theta^k=  U* \phi 
\nonumber \\
\phi_k & \rightarrow & {\phi'}_k
=\sum_{0\le l \le m \le p \le k}
\partial^{m-l}\bar{\partial}^l U_{p-m}
\bar{\partial}^{m-l} \partial^l \phi_{k-l} \frac{(-1)^l}{2^m l!(m-l)!} \ .
\label{noncomU(1)}
\end{eqnarray}
Note that for $\phi_0$ the gauge transformation is the same as the
commutative U(1) theory.

%%%%%%%%%%%%%%%%%%%%%%%%%%%%%%%%%%%%%%%%%%%%%%%%%%%%%
%Then $\partial_{\mu} {\phi}$ is not covariant 
%under this gauge transformation.
%Instead of $\partial_{\mu} \hat{\phi}$,
Let us define the covariant derivative operator by
\begin{equation}
 \nabla_{\mu}=\partial_{\mu}-i A_{\mu}\;,\label{EQ*:covariant-derivative}
\end{equation}
where $A_{\mu}$ is a local 1-form %(section of tangent fiber on ${\bb R}^2$)
whose gauge transformation is defined by 
\begin{equation}
\textsf{A} \rightarrow  i g * \textsf{d} * g^{-1} + g * \textsf{A} * g^{-1} .
\end{equation}
{}From this gauge transformation,
we find that
\begin{equation}
 {\nabla}_{\mu} * {\phi}
  :={\partial}_{\mu} {\phi}- i{A}_{\mu} * {\phi}.
\end{equation}
is covariant under gauge transformation.

In the complex coordinates $
  A
  =\frac{1}{\sqrt{2}}(A_1-iA_2) $ and $
  \bar{A}
  =\frac{1}{\sqrt{2}}(A_1+iA_2)$,
the gauge transformations are 
\begin{eqnarray}
A \rightarrow i g * \partial g^{-1} + g * A * g^{-1}
\ , \ \ \ 
\bar{A} \rightarrow -i \bar{\partial}g  * g^{-1} + 
g * \bar{A} * g^{-1}
\end{eqnarray}
The curvature components of the connection $A$ are given by
\begin{eqnarray}
 F_{z z}&=&F_{\bar{z} \bar{z}}=0\nonumber\\
 F_{z \bar{z}}&=&iF_{12}=\partial_{z}A_{\bar{z}}-\partial_{\bar{z}}A_{z}
  -i[A_{z},A_{\bar{z}}]_* \ , \nonumber
\end{eqnarray}
where $[A , B]_* := A*B - B*A$ .
%If we consider 2+1 dimension case, static energy of the 
%gauge field is described by
The magnetic field (in the sence of 2+1 dimension model) is defined by
\begin{equation}
B := -i F_{z \bar{z}}\ \ .
\end{equation}
Although we are using the same notation for the curvature as for the 
commutative ${\mathbb R}^2$, in the following, 
we consider only the noncommutative ${\mathbb R}^2$
so the notation should be clear.
%We use common notation of the curvature with commutative ${\mathbb R}^2$.
%In the following, we consider only noncommutative ${\mathbb R}^2$
%so we might not be confused by these notation.

Using these complex coordinates, the covariant derivatives of the Higgs fields are
\begin{eqnarray}
D * \phi = ( \partial -i A ) * \phi  &,& \ \ \
\bar{D}* \phi = (\bar{\partial} - i \bar{A} ) * \phi \ , \\
D * \bar{\phi}  = \partial \bar{\phi} + i \bar{\phi} * A  &,& \ \ \
\bar{D} * \bar{\phi} = \bar{\partial}\bar{\phi} + i \bar{\phi} * \bar{A}
\ .
\end{eqnarray}

%%%%%%%%%%%%%%%%%%%%%%%%%%%%%%%%%%%%%%%%%%%%%%%%%%%%%%%%%%%%%%%%%%%%%
%%%%%%%%%%%%%%%%%%%%%%%%%%%%%%%%%%%%%%%%%%%%%%%%%%%%%%%%%%%%%%%%%%%%%
\subsection{The Action Functional and The Vortex Equations}
The action functional for the noncommutative Abelian
Higgs model \cite{Ginzburg_Landau} is given by
\begin{eqnarray}
S= \int d^2 z \ \Big\{ -\frac{1}{2}  (F_{z \bar{z}})^2_* 
+ D * \phi * \bar{D} * \bar{\phi} + \bar{D} * \phi * D * \bar{\phi}
+ \frac{1}{2} (\phi * \bar{\phi} -1 )^2 \Big\} \ .
\nonumber \\
\end{eqnarray}
As in the commutative case, $S$ can be rewritten as
\begin{eqnarray}
S = S_T 
+ \int d^2 z \ \left\{ 2 \bar{D} * \phi * D * \bar{\phi} 
+ \frac{1}{2} ( B+ (\phi * \bar{\phi} -1 ))_*^2 \right\}
\ \ , \\
S_T := \int  \ \Big[ \frac{1}{2} \left\{
{\textsf d} ( i \phi * {\textsf d_A} * \bar{\phi} 
- i ({\textsf d_A} * \phi )* \bar{\phi} )
\right\} + {\textsf B} \ \Big] \ \ .
\end{eqnarray}
$S_T$ is a topological term.

Therefore the vortex equations are given by
\begin{eqnarray}
\bar{D} * \phi = ( \bar{\partial} - i \bar{A} )* \phi = 0 
\ \ , \ \ 
B+ \phi * \bar{\phi} -1 = 0 \ \ . \label{BPS}
\end{eqnarray}
We call solutions of these equations noncommutative vortices or noncommutative
vortex solutions.
Some solutions in 
\cite{Bak,Bak_Lee_Park,Jatkar_Mandal_Wadia,Lozano_Moreno_Schaposnik,Hamanaka_Terashima,Popov,add1,add2} have been constructed 
by using the operator formalism. 
These are different from the solutions 
discussed in section \ref{section vortex}.
%(see also section \ref{another type}).

The formal expansions of the fields are
\begin{equation}
\phi = \sum_{n=0}^{\infty} \theta^n \phi_n (z , \bar{z} )
\ , \ \ 
A = \sum_{n=0}^{\infty} \theta^n A_n (z , \bar{z} )
\  .
\end{equation}
The $k$-th order equations for (\ref{BPS}) are
\begin{eqnarray}
-i(\partial \bar{A}_k + \bar{\partial} A_k) + 
\phi_k \bar{\phi}_0 + \phi_0 \bar{\phi}_k - \delta_{k 0} + 
C_k(z, \bar{z} ) &=& 0  \label{BPS1}\\
\bar{\partial} \phi_k - i \bar{A}_k \phi_0 
-i\bar{A}_0 \phi_k + D_k(z, \bar{z} ) &=& 0 . \label{BPS2}
\end{eqnarray}
Here $C_k(z, \bar{z} )$ is the coefficient of
$\theta^k$ in $-[ A , \bar{A}]_* + \phi * \bar{\phi}- 
( \phi_k \bar{\phi}_0 + \phi_0 \bar{\phi}_k ) ,$
so
$C_k(z, \bar{z} )$ is a function of 
$\{ A_i , \bar{A}_j , \phi_m , \bar{\phi}_n | 
0 \le i , j , m , n \le k-1 \}$.
Similarly, 
$D_k(z, \bar{z} )$ is the coefficient of $\theta^k$ in
$ -i \bar{A} * \phi - (- i \bar{A}_k \phi_0 
-i\bar{A}_0 \phi_k )$ 
and a function of 
$\{ A_i , \bar{A}_j , \phi_m , \bar{\phi}_n | 
0 \le i , j , m , n \le k-1 \}$.

In particular in the case of $k=0$, 
(\ref{BPS1}) and (\ref{BPS2}) coincide with 
the commutative U(1) vortex
equations (\ref{BPS_1})
i.e.,
$
\bar{D} \phi_0 = ( \bar{\partial} - i \bar{A}_0 ) \phi_0 = 0 
$ and $
B_0+ \phi_0  \bar{\phi}_0 -1 = 0 $,
where $B_0 = -i (\partial \bar{A}_0 - \bar{\partial} {A}_0 )$.

In the region $\phi_0 \neq 0$, substituting (\ref{BPS2}) 
into (\ref{BPS1}) for $A_k$ and $\bar{A}_k$, we get 
\begin{eqnarray}
&&  \!\!\!\!\!\!\!\!\! \left\{  
\frac{\partial \phi_0}{\phi_0^2}
(\bar{\partial} \phi_k -i \bar{A}_0 \phi_k +D_k )
-\frac{1}{\phi_0}
(\Delta  \phi_k -i \partial \bar{A}_0 \phi_k 
 -i \bar{A}_0 \partial \phi_k + \partial D_k ) \right\}
+ \{ c.c. \} \nonumber \\
&& +\phi_k \bar{\phi}_0 + \phi_0 \bar{\phi}_0 -\delta_{k0} + C_k = 0.
\label{BPS3}
\end{eqnarray}
Here $\{ c.c. \}$ is the complex conjugate of preceding terms
and $\Delta = \partial \bar{\partial}$.

%%%%%%%%%%%%%%%%%%  Ver2
Setting
\begin{equation}
\varphi_k := \frac{\phi_k}{\phi_0}+\frac{\bar \phi_k}{\bar \phi_0}=
2 Re \big( \frac{\phi_k}{\phi_0}\big)\ \ \mbox{and} \ \ 
d_k = \frac{D_k}{\phi_0} \ ,
\end{equation}
by (\ref{BPS3}), $\varphi_k$ , $d_k$ satisfy
\begin{equation}
(- \Delta + |\phi_0|^2 ) \varphi_k = E_k \, \label{BPS4}
\end{equation}
where
\begin{equation}
E_k:= -C_k + \partial d_k - \bar{\partial} \bar{d}_k .
\end{equation}
{}From (\ref{taubes2}), there exists a positive constant
$C$ such that
\begin{equation}
|D_1 | < C \frac{1}{1+r^3} \ , \ 
|C_1 | < C\frac{1}{1+r^4} \ , \ 
|E_1 | < C\frac{1}{1+r^4}. \label{CDEasympt}
\end{equation}
We will use (\ref{CDEasympt}) to prove some of our main theorems.
But in the proofs the actual power of $r$ is not important
\footnote{
{}From a naive observation, we get $|D_1 | < C \frac{1}{1+r^3} \ , \ 
|C_1 | < C\frac{1}{1+r^2} \ , \ 
|E_1 | < C\frac{1}{1+r^2}$. 
But we can constrain fields by choosing a gauge condition.
For example, from the gauge condition $Im\; \phi_0 =0$,
we can derive (\ref{CDEasympt}).
The following discussion holds for both cases.
}.

%%%%%%%%%%%%%%%%%%%%%%%%%%%%%%%%%%%%%%%%%%%%%%%%%%%%%%%%%%%%%
\subsection{Preliminary Facts}

As in section \ref{sect2}, 
the vortex number $N_0 := \frac{1}{2} \int B_0 $
is a integer corresponding to the 
winding number of $\lim_{|z| \rightarrow \infty} \phi_0 $.

Let $(A_0 , \phi_0 )$ be a smooth solution of (\ref{BPS_1}).
Define $I_k$ and $w(\bar{z})$ by
\begin{eqnarray}
I_k(z , \bar{z}) = \exp \left( \int \frac{1}{\phi_k} (D_k - i \bar{A}_k \phi_0 ) \right) \ \ , 
\ \ 
w(\bar{z}) = \frac{1}{2\pi} \int_{\cal B}
\frac{\bar{A}_0}{\zeta - \bar{z}} d\zeta \wedge d\bar{\zeta} \ ,
\nonumber \\
\end{eqnarray}
where ${\cal B}$ is a closed disc in ${\mathbb C}$.
Using $I_k$ and $w(\bar{z})$, the following theorem is given as well as
the Theorem \ref{Taubes-1}. %%%% 
\begin{thm} 
Let $\{ (A_i , \phi_i ) \ ; \  0 \le i \le k\}$ be a smooth solution of (\ref{BPS}).
Then $\left(e^{-w} I_k(z,\bar{z}) \phi_k(z , \bar{z} ) \right)$
is complex analytic, that is 
\begin{equation}
\bar{\partial} \left(e^{-w} I_k(z,\bar{z}) \phi_k(z , \bar{z} ) \right)=0 .
\label{ana}
\end{equation}
\end{thm}

\noindent
[Proof]\  We note that
\begin{equation}
\bar{\partial} \left(e^{-w} I_k(z,\bar{z}) \phi_k (z , \bar{z} ) \right)
=e^{-w} ( (\bar{\partial}I_k)\phi_k
+I_k \bar{\partial} \phi_k - (\bar{\partial}w) I_k \phi_k ). \label{p_1}
\end{equation}
By definition, 
\begin{equation}
\bar{\partial} I_k = \left( \frac{D_k}{\phi_k} - i \bar{A}_k \frac{\phi_0}{\phi_k}  \right) I_k
\ \ , \ \ 
\bar{\partial} w = i \bar{A}_0 \ .  \label{p_2}
\end{equation}
{}From 
(\ref{p_1}) and (\ref{p_2}), we get (\ref{ana}). 
\begin{flushright}
$\square $
\end{flushright}

Note that $e^{-w}$ is a non-vanishing function.
The holomorphic function $\Omega (z) := e^{-w} I \phi_k$
has a finite number of zeros in any bounded set ${\cal B}$.
In a neighborhood of each zero $z_a$, there is a nonvanishing function
such that 
$\Omega (z) = (z- z_a)^{n_a} \Omega_a (z)$.
\\

\begin{thm}
Let $\{ (A_i , \phi_i ) \ ; \  0 \le i \le k\}$ 
be a smooth, locally $L^2$ solution of (\ref{BPS}).
%Then $\{ (A_i , \phi_i ) \ ; \  0 \le i \le k\}$  
%is smooth and 
There exist $N$ points $\{ z_1 , \dots , z_N \} $ in 
$\mathbb{C} $, such that
\begin{equation}
Z(\phi_k):= \{ z \in \mathbb{C} : I_k \phi_k (z) =0 \}=\{ z_1 , \dots , z_N \} .\end{equation}
There is a neighborhood of each $z_a$ in which
\begin{equation}
I_k \phi_k (z) = (z- z_a)^{n_a} h_a (z) \ ,
\end{equation}
where $n_a$ is the multiplicity of the 
point $z_a$ in $\{ z_1 , \dots , z_N \}$ , and
$ h_a (z)= e^w \Omega_a$ is a $C^\infty $, nonvanishing function.
\end{thm}

%%%%%%%%%%%%%%%%%%%%%%%%%%%%%%%%%%%%%%%%%%%%%%%%%%%%%%%%%%%%%%%%%%%%%%%%%%%
%%%%%%%%%%%%%%%%%%%%%%%%%%%%%%%%%%%%%%%%%%%%%%%%%%%%%%%%%%%%%%%%%%%%%%%%%%%
\section{Vortex Number} \label{section vortex}
In this section, we show that the vortex number is constant
for vortex solutions that are given by noncommutative deformations of
Taubes' vortex solutions.
%%%%%%%%%%%%%%%%%%%%%%%%%%%%%%%%%%%%%%%%%%%%%%%%%%
\subsection{Noncommutative Vortex Number}
We first study conditions which preserve the vortex number
under a noncommutative deformation.
\begin{thm}\label{vortex number1}
If the vortex number of a classical solution (\ref{BPS_1}) is 
$\frac{1}{2\pi}\int d^2 x B_0 =N_0$ and 
$|\phi_k| < Cr^{-\epsilon} $,
$|\partial_r \phi_k |< Cr^{-\epsilon+1}$, for some $\epsilon > 0$ and large $r$, then
\begin{equation}
\frac{1}{2 \pi} \int d^2 x B = N_0 \ .
\end{equation}
\end{thm}

[Proof] \ 
Let $F_k$ be the coefficient of $\theta^k$ in $F_{12}$.
Then we have for $k >0$
\begin{eqnarray}
\int d^2x \ F_k &=& -i \int d^2x\   (\partial \bar{A}_k - \bar{\partial} A_k ) 
- [A , \bar{A}]_* |_k \nonumber \\
&=& \oint \textsf{A}_k  \nonumber \\
&- &\hspace{-8mm} \int 
\sum_{l+m+n=k , n\ge 1} \Big\{
A_l
(\overleftarrow{\partial} \frac{1}{2} \overrightarrow{\bar{\partial}}
-\overleftarrow{\bar{\partial}} \frac{1}{2}\overrightarrow{\partial})^n
\frac{1}{n!}
\bar{A}_m
- \bar{A}_m
(\overleftarrow{\partial} \frac{1}{2} \overrightarrow{\bar{\partial}}
-\overleftarrow{\bar{\partial}} \frac{1}{2}\overrightarrow{\partial})^n
\frac{1}{n!}
A_l
\Big\}
\nonumber \\
&=& \oint \frac{1}{i \phi_0} 
(\bar{\partial} \phi_k - \bar{A}_0 \phi_k + D_k ) +c.c. 
\nonumber \\
&&- i \oint 
\sum_{l+m+n=k,n\ge1}\!\!\!\!\!\! \Big\{
A_l
(\overleftarrow{\partial} \frac{1}{2} \overrightarrow{\bar{\partial}}
-
\overleftarrow{\bar{\partial}} \frac{1}{2}\overrightarrow{\partial})^{(n-1)}
\frac{1}{n!}
\textsf{d} \bar{A}_m
\nonumber \\ 
&&\mbox{}\hspace{3cm}
- \bar{A}_m
(\overleftarrow{\partial} \frac{1}{2} \overrightarrow{\bar{\partial}}
-
\overleftarrow{\bar{\partial}} \frac{1}{2}\overrightarrow{\partial})^{(n-1)}
\frac{1}{n!}
\textsf{d} A_l
\Big\} \ .
\end{eqnarray}
{}From the following facts, we get the result that we want.
\begin{eqnarray}
|D_k | \le C  \frac{1}{r^{2+\epsilon}} \ , \ 
| \partial \bar{A}_{k-1} \partial \phi_0 |
\le \frac{1}{r^{2+\epsilon}} \\
| \bar{A}_0 \phi_k | \le C \frac{1}{r^{1+\epsilon}} \\
| \bar{\partial}\phi_k | \le C \frac{1}{r^{1+\epsilon}} \ , \ 
A_k \le C \frac{1}{r^{1+\epsilon}} ,
\end{eqnarray}
where $C$ is a constant. 
We use (\ref{BPS1}), (\ref{BPS2}) and (\ref{CDEasympt}) here.
Then, $\int d^2x F_k  =0$.
\begin{flushright}
$\square $
\end{flushright}

%%%%%%%%%%%%%%%%%%%%%%%%%%%%%%%%%%%
%As we saw in (\ref{noncomU(1)}),
%the gauge transformation for general $\phi_k$ is complex, but 
%for $\phi_0$ gauge transformation is same as commutative one 
%(see section \ref{ncU(1)}).
%Therefore, we use it to simplify the equations
%(\ref{BPS1}) and (\ref{BPS2}).
%After demanding the condition 
%
%\begin{eqnarray}
%\phi_0 = \bar{\phi}_0 , 
%\end{eqnarray}
%then the (\ref{BPS3}) is represented as
%\begin{eqnarray}
%\left\{ (\bar{\partial}-i\bar{A}_0)(\partial + i A_0) -\phi_0^2 
%\right\} \phi_{R,k} = \frac{-1}{2}E_k \ , \label{BPS4}
%\end{eqnarray}
%where $\phi_{R,k}$ is real part of $\phi_k$ and 
%the real function $E_k$ is defined by
%\begin{eqnarray}
%E_k := \left( \partial D_k - \frac{\partial \phi_0}{\phi_0}D_k \right) 
%+(c.c) + \phi_0 C_k \ .
%\end{eqnarray}

%\begin{thm}
%The solution of (\ref{BPS4}) exist in the domain that
%real function $\phi_0$ satisfies 
%\begin{eqnarray}
%\frac{1}{2} + \frac{|\partial \phi_0|^2 }{\phi_0^2}
%\le \frac{3}{2} \phi_0^2
%\end{eqnarray}
%\end{thm} 
%

We next show the following theorem.
\begin{thm} \label{orderofADCE}
Let $\phi_k , A_k , D_k , C_k , E_k$ be fields and functionals
defined above.
$\phi_k =O({r^{-\alpha_k}})$,
$A_k = O({r^{-\beta_k}})$,
$D_k = O({r^{-\delta_k}})$,
$C_k = O({r^{-\gamma_k}})$ and
$E_k = O({r^{-\eta_k}})$,
where
$\alpha_k = 2k$,
$\beta_k = 2k + 1$,
$\gamma_k = 2k+2$,
$\delta_k = 2k+1$ and
%%%%%%%%% Ver2
$\eta_k = 2k+2$ for $ k \in {\mathbb Z}_{>0}$
\footnote{
Note that without the gauge fixing condition 
$Im\; \phi_0 =0$, we can easily derive
$\gamma_k = 2k$ and $\eta_k = 2k+2$.}.
\end{thm}

[Proof] \
The proof is by induction.\\
(I) 
{}From asymptotic behaviors (\ref{asym_phi0}) and (\ref{asym_A0}) and 
the vortex equations (\ref{BPS1}) and (\ref{BPS2}),
for $k=1$ we get 
$\alpha_1 = 2$, $\beta_1 = 3$, $\gamma_1 = 2$,
$\delta_1= 3$ and $\eta_1 = 2$. \\
%(I) For $k=2$, using (\ref{BPS1}) and (\ref{BPS2}),
%$\alpha_2 = 4$, $\beta_2 = 5$, $\gamma_2 = 4$,
%$\delta_2= 5$ and $\eta_2 = 4$.\\
(II)
Assume above the theorem for $k=1, \dots , j-1$. 
{}By the definition of $D_k$, there exists a positive constant $C$ such that
\begin{equation}
| D_j | < C \left\{ \sum_{i=1}^{j-1} \frac{1}{r^{(\alpha_{j-i}+\beta_i )}} +
\sum_{n=1}^{j}\sum_{i=0}^{j-n} 
\frac{1}{r^{(\alpha_{j-i-n}+\beta_i +2n)}} \right\}
= O\big( \frac{1}{r^{2j+1}} \big) \ .
% \ , \nonumber \\
%D_{j-1} \sim \sum_{i=1}^{j-2} \frac{1}{r^{(\alpha_{j-i-1}+\beta_i )}} +
%\sum_{n=1}^{j-1}\sum_{i=0}^{j-n-1} 
%\frac{1}{r^{(\alpha_{j-i-n-1}+\beta_i +2n)}} \ ,
\end{equation}
%For $i=0, \dots ,j-n-1$, $n=1, \dots , j-1$,
%the relation of each power of the second terms of $D_j$
%and all terms of $D_{j-1}$ are
%\begin{eqnarray}
%\alpha_{j-i-n}+\beta_i +2n = 
%(\alpha_{j-i-n-1}+\beta_i +2n)+2 \ , \label{d-1}
%\end{eqnarray}
%from induction hypothesis.
%For $n=0$ in $D_j$ ,
%i.e. $\sum_{i=1}^{j-1} \frac{1}{r^{(\alpha_{j-i}+\beta_i )}} $, 
%there is no corresponding terms is $D_{j-1}$.
%{}From induction hypothesis,
%\begin{eqnarray}
%\alpha_{j-i}+\beta_i = \alpha_0+\beta_0 +2j  \ . \label{d-2}
%\end{eqnarray}
%{}From (\ref{d-1}) and (\ref{d-2}), we get 
%$\delta_j=\delta_{j-1}+2$ .
Therefore, $\delta_j = 2j+1$.
With this result for $\delta_j$,
we can prove the statements for $\alpha_k $,
$\beta_k $,
$\gamma_k $ and
$\eta_k $, by similar arguments.
\begin{flushright}
$\square $
\end{flushright}

%Using these facts, we found that 
%\begin{eqnarray}
%D_k \sim \frac{1}{r^{1+2k}} \ , \ 
%C_k \sim \frac{1}{r^{2+2k}} \ , \ 
%E_k \sim \frac{1}{r^{2+2k}} .
%\end{eqnarray}

%%%%%%%%%%%%%%%%%%%%%%%%%%%%%%%%%%%%%%%%%%
%%%%%%% Ver 2
%%%%%%%%%%%%%%%%%%%%%%%%%%%%%%%%%%%%%%%%%%

%%%%%%%%%%%%%%%%%%%%%%%%%%%%%%%%%%%%%%%%%%%%%%%%%%%%%%%
\subsection{The Schr\"odinger equation and Vortex Solutions}
To show that there exists a unique noncommutative vortex solution
deformed from the Taubes' vortex solution, 
we consider the stationary Schr\"odinger equation
\begin{equation}
(-\Delta + V(x) )u(x)=f(x)\  \label{(1)}
\end{equation}
in ${\mathbb R}^2$, where $V(x)$ is a real valued $C^{\infty}$ function.
%Let us introduce two differential operator 
%\begin{eqnarray}
%P_0\ u(x) &=& f(x) \  ,  \ P_0 := (-\Delta + 1 ) \ , \label{P_0} \\
%P_1\ v(x) &=& g(x) \  ,  P_1 := (-\Delta + V(x) ) \ , \ \label{P_1}
%\end{eqnarray}
Throughout this section, we impose the 
following assumptions for $V(x)$
%%%%%%%%%%%%%%%%%%%%%%%%%% V assum %%%%%%%%%%%%%%%%%
\begin{eqnarray}
&(a1)& V(x) \ge 0 \ , \ {}^{\forall} 
x \subset {\mathbb R}^2 \label{a1} \\
&(a2)& \mbox{There exist } 
K \subset  {\mathbb R}^2 \ \mbox{and}\ {}^{\exists} c >0 
\mbox{ such that $K$ is a compact set and} \nonumber\\
&&\mbox{for}\  x \in {\mathbb R}^2  
\backslash K \ , \ 
 V(x) \ge c   \label{a2}\\
&(a3)& \mbox{There exist } x_1, \dots , x_N \in {\mathbb R}^2  \mbox{ such that }\ V(x_i) =0 , V(x)>0 \ \nonumber \\
&& \mbox{for}\ x \not{\!\in} 
\{ x_1 , \dots , x_N \}  %\nonumber \\
 \label{a3} \\
&(a4)& \mbox{For any}\ \alpha = (\alpha_1 , \alpha_2)\in {\mathbb Z}_+^2,
 \ \mbox{There exists a positive constant} \ C_{\alpha}\ \nonumber \\
&& \mbox{such that }\
 |\partial_x^{\alpha} (V- c) | \le C_{\alpha} \label{a4}
 \ \mbox{for any $x \in {\mathbb R}^2$} \label{a4}
\end{eqnarray}
%%%%%%%%%%%%%%%%%%%%%%%%%%%%%%%%%%%%%%%%%%%%%%%%%%%%%
We note that the system (\ref{BPS4}) satisfies
the assumptions $(a1)-(a4)$. 
We set
\begin{equation}
H_l(n):= \{ f | 
\ \ || f || := \sup_{x \in {\mathbb R}^2} (1+ |x|^n)|
\partial_x^{\alpha} f(x)| < \infty \ \mbox{for any}\ |\alpha|\le l \}
\label{E}
\end{equation}
for $n \in {\mathbb Z}_+$.
We let $C ,C_\alpha,$ etc. denote unimportant
positive constants whose value may change from line to line unless otherwise
stated.
The next theorem's proof follows a series of lemmas.

%%%%%%%%%%%%%%%%%%%%%%% 
\begin{thm}\label{finitetheo}
Under the assumptions $(a1)-(a4)$,
there exists a unique solution $u \in H_l(n)$ of (\ref{(1)}) for 
any $f \in H_l(n)$. %then $\exists u  $.
\end{thm}

Following Theorem 2.1 (iii), Theorem 3.3 and Theorem 3.8 in 
\cite{Pinsky}, we have
\begin{lem}\label{prop0}
Under the  assumptions $(a1)-(a4)$, $V$ is subcritical, i.e.
There exists a positive solution $G(x,y)$ of
\begin{equation}
(-\Delta +V(x))  G(x,y) = \delta^2(x-y)\ . \label{green_add}
\end{equation}
\end{lem}
Consider the stationary Schr\"odinger
equation
\begin{equation}
(-\Delta + c) u(x) =f(x)
\end{equation}
in ${\mathbb R}^2$, where $c$ is a positive constant.% in $(a2)$.
The Green's function $G_c(x,y)$ for $(-\Delta + c)$
is given explicitly by
\begin{equation}
G_c(x,y)=\frac{1}{2\pi} K_0(\sqrt{c}|x-y|) 
=\int_0^{\infty} \frac{\cos(\sqrt{c}|x-y|t)}{\sqrt{t^2+1}} dt
%=\sqrt{\frac{\pi}{2|x-y|}}\frac{e^{-|x-y|}}{\Gamma (1/2)}
%\int_0^\infty e^{-t} t^{-1/2}\left( 1+\frac{t}{2|x-y|} \right)^{-1/2} dt
\label{green_function}
\end{equation}
where $K_0(z)$ is the modified Bessel function,
with known asymptotic behavior (cf.\cite{Pinsky})
\begin{eqnarray}
K_0 (z) &\sim & \sqrt{\frac{\pi}{2z}}e^{-z} 
\ \ \mbox{for}\  |z| \gg 1
\nonumber \\
K_0 (z) &\sim & \log |z| \ \ \mbox{for}\  0< |z| \ll 1  \ .
\label{asym_K0}
\end{eqnarray}

Let us estimate the behavior of the Green's functions in 
(\ref{green_add})
at large and small $|x-y|$.
%We propose the following 
%%%%%%%%%%% Prop.1
\begin{lem}\label{prop1}
Assume $(a1)-(a4)$. For $|x-y| \le r_0 \  
(0 < r_0 <1)$, there exist constants $C_1$ and $C_2$
such that
\begin{equation}
-C_1 \log|x-y| \le G(x,y) \le -C_2 \log|x-y| \ .
\end{equation}
\end{lem}
The proof of this lemma is given in \cite{Miranda}
(cf. Theorem 4.2 in \cite{Pinsky}).

%%%%%%%%%%% Prop.2
\begin{lem}\label{prop2}
Assume $(a1)-(a4)$. 
Let $r_1$ be the radius
of a disk ${\cal D}_1$ 
centered at the origin and
with $K \subset {\cal D}_1$.
For $|x-y| \ge r_1 $, 
there exists a constant $C$
such that
\begin{equation}
G(x,y) \le C G_c(x,y) \ .
\end{equation}
\end{lem}
%%%%%%%%%%%%%%%%%%%%%
[Proof]\
For $|x| \ge r_1 $,
\begin{equation}
(-\Delta + V(x))G_c(x,y) =(V(x)-c) G_c(x,y) \ge 0,
\end{equation}
where we use (\ref{asym_K0}). 
Therefore $G_c(x,y)$ is a superharmonic function
with respect to the $(-\Delta + V(x))$.
Since ${\cal B}_1 = \partial {\cal D}_1$ is compact,
there exists a positive constant $C$ such that
\begin{equation}
G(x,y) \le C G_c(x,y) \ \mbox{for}\  \ x \in {\cal B}_1 \ .
\end{equation}
By the maximal principle we get 
\begin{equation}
G(x,y) \le C G_c(x,y) \ \mbox{for}\  \ |x-y| \ge r_1 \ .
\end{equation}
\begin{flushright}
$\square$
\end{flushright}

Now, using Lemmas \ref{prop0}-\ref{prop2}, we show Theorem \ref{finitetheo}.

[Proof of Theorem \ref{finitetheo}]\\
To show $u \in H_l(n)$, we estimate $(1+|x|^n) u(x)$. 
It is enough to consider the case 
$|x| \ge r_0$ or the fixed $r_0$.
%In the following, we consider only the case of 
%$|x| \ge r_0$ for the fixed $r_0$.
In this case,
\begin{eqnarray}
(1+|x|^n) u(x) &=& \int (1+|x|^n) G(x,y) f(y) dy \nonumber \\
&=& \int_{|x-y|\le r_0} (1+|x|^n) G(x,y) f(y) dy  \label{case1} \\
&&+ \int_{|x-y|\ge r_1} (1+|x|^n) G(x,y) f(y) dy  \label{case2} \\
&&+ \int_{ r_0 \le |x-y|\le r_1} (1+|x|^n) G(x,y) f(y) dy \label{case3}
\end{eqnarray}

\noindent
(I) Estimation of (\ref{case1}) %(Case of $|x-y| \le r_0$) \\
\begin{eqnarray}
(\ref{case1})
&\le & (1+|x|^n) 
 \left| \int_{|x-y|\le r_0} G(x,y) f(y) dy \right|
\nonumber \\
&\le &  C   \int_{|x-y|\le r_0}  | G(x,y) |\ | (1+|y|^4) f(y) |   dy  
\nonumber \\
&\le & C'' \int_0^{r_0} r \log{r} dr = C''' \label{case1result}
\end{eqnarray}
Here we use the facts that there exists some constant $C$ such that 
$1+|x|^4 < C (1+|y|^4)$ and we use Lemma \ref{prop1}.\\

\noindent
(II) Estimation of (\ref{case2}) %(Case of $|x-y| \ge r_1$) \\
\begin{eqnarray}
(\ref{case2})
&\le& C  \int_{|x-y|\ge r_1}\frac{1}{2\pi}\sqrt{\frac{\pi}{2\sqrt{c}|x-y|}}
 e^{-\sqrt{c}|x-y|}(1+|y|^n)^{-1} (1+|y|^n) |f(y)| dy  \nonumber \\
&\le& C' \int_{|x-y|\ge r_1}\frac{1}{2\pi}\sqrt{\frac{\pi}{2\sqrt{c}|x-y|}}
 e^{-\sqrt{c}|x-y|}(1+|y|^n)^{-1} dy.\   
 \label{II_1}
\end{eqnarray}
Here we use (\ref{asym_K0}).
Let us introduce two subregions 
${\cal A}_1(x,r_1,r_2)= \{ y \in {\mathbb R}^2 \ |\ 
|x-y|\ge r_1, \ |y| \le r_2 , \  
\mbox{for fixed}\ x \}$
and ${\cal A}_2(x,r_1,r_2)= \{ y \in {\mathbb R}^2 \ |\ 
|x-y|\ge r_1 , \ |y| \ge r_2  , \  
\mbox{for fixed}\ x \}$.
\begin{eqnarray}
(\ref{II_1})= C' (\int_{{\cal A}_1} + \int_{{\cal A}_2} )
 \frac{1}{2\pi}\sqrt{\frac{\pi}{2\sqrt{c}|x-y|}}
 e^{-\sqrt{c}|x-y|}(1+|y|^n)^{-1}  dy  \ .\ \label{II_2} 
\end{eqnarray}
We estimate the first term of (\ref{II_2}).
\begin{eqnarray}
\int_{{\cal A}_1}\frac{1}{2\pi}\sqrt{\frac{\pi}{2\sqrt{c}|x-y|}}
 e^{-\sqrt{c}|x-y|}(1+|y|^n)^{-1} dy 
 \nonumber \\
 \le C \int_{r_1}^{\infty}
 \frac{1}{\sqrt{c^{1/2}r}}(1+r^n) e^{-\sqrt{c}r} r \ dr  \le C' \ .
 \label{II_3}
\end{eqnarray}
Next we estimate the second term of (\ref{II_2}).
\begin{eqnarray}
\int_{{\cal A}_2}\frac{1}{2\pi}\sqrt{\frac{\pi}{2\sqrt{c}|x-y|}}
 e^{-\sqrt{c}|x-y|}(1+|y|^n)^{-1} dy 
 \nonumber \\
\le C\int_{{\cal A}_2} \sqrt{\frac{1}{\sqrt{c}|x-y|}}
 e^{-\sqrt{c}|x-y|}\big(1+\frac{|x-y|^n}{1+|y|^n} \big) dy \nonumber \\
\le C' \int_{r_1}^{\infty}
 \frac{1}{\sqrt{c^{1/2}r}}(1+r^n) e^{-\sqrt{c}r} r \ dr  \le C'' \label{II_4}
\end{eqnarray}
(\ref{II_3}) and (\ref{II_4}) show that $(\ref{case2})< C$. \\

\noindent
(III) Existence of some constant $C$ such that $(\ref{case3}) < C$
is trivial because the region of integration in (\ref{case3}) is compact.\\

Differentiating (\ref{(1)}) sufficiently and using
similar computations as above, we obtain the estimate for 
$(1+|x|^n) | \partial^{\alpha}_x u| < \infty \ (|\alpha| \le l )$.\\

\noindent
{}From (I)-(III), we have Theorem \ref{finitetheo}.
\begin{flushright}
$\square$
\end{flushright}

%%%%%%%%%%%%%%%%%%%%%%%%%%%%%%%%%%%%%%%%%%%%%%%%%%%%%%%%

Equation (\ref{BPS4}) is a particular example of (\ref{(1)}), so
Theorem \ref{vortex number1} and \ref{finitetheo} imply 
the following theorem.
%%%%%%%%%%%%%%%%
\begin{thm}
Let $A_0$ and $\phi_0$ be a Taubes' vortex solution stated in section \ref{sect2}, in other words, 
$(A_0 , \phi_0)$
satisfy the equations (\ref{BPS_1}) with the condition (\ref{taubes2}).
Then there exists a unique solution $( A , \phi )$ of the noncommutative vortex
equations (\ref{BPS})
with
$A|_{\theta =0 }=A_0 ,\ \phi |_{\theta=0}=\phi_0$, and its vortex
number is preserved:
\begin{equation}
N=N_0 \ , \  \mbox{i.e.} \ \ \frac{1}{2\pi} \int d^2x \; B = 
\frac{1}{2\pi} \int d^2x \; B_0 \ .
\end{equation}
\end{thm} 

[Proof]
Consider (\ref{(1)}) with $V(x)=|\phi_0|^2$ and $f(x)=E_k$ .
{}From the facts in section \ref{sect2}, we find $V(x)$ satisfies $(a1)-(a4)$.
Next, we consider $E_k$.
{}From (\ref{CDEasympt}), $E_1 \in H_{\infty}(4)$. 
If $ E_i \in H_{\infty}(2i+2) (i= 1, \dots , k-1)$, as a result of Theorem \ref{finitetheo},
there exist unique solutions $\varphi_1 , \dots , \varphi_{k-1}$.
Then we find $E_k \in H_{\infty}(2k+2)$ from Theorem \ref{orderofADCE}.
Therefore $E_k \in H_{\infty}(2k+2)$ is proved for arbitrary $k$.
Theorem \ref{finitetheo} is applicable to (\ref{BPS4}) for arbitrary $k$, then it is shown that each $\varphi_k $ is determined uniquely. 
Finally, Theorem \ref{orderofADCE} and Theorem \ref{vortex number1} imply that
$N=N_0$.
\begin{flushright}
$\square$
\end{flushright}

%%%%%%%%%%%%%%%%%%%%%%%%%%%%%%%%%%%%%%%%%%%%%%%%%%%%%%%%%%%%%
%%%%%%%%%%%%%%%%%%%%%%%%%%%%%%%%%%%%%%%%%%
\section{Noncommutative Vortex Solutions via the Fock Representation}
\label{anothe type}
Solutions of (\ref{BPS}) are given in
\cite{Bak,Bak_Lee_Park,Jatkar_Mandal_Wadia,Lozano_Moreno_Schaposnik,Hamanaka_Terashima,Popov}, etc.
These solutions are substantially different from the solution discussed 
in the previous section. The difference will be clear soon.
In this section, we show the existence of bounded solutions 
via Fock space formalism.
As a simple example, we investigate the properties of 
the solution in \cite{Bak}.

%%%%%%%%%%%%%%%%%%%%%%%%%%%%%%%%%%%%%%%%%%%%%%%%%%%%%%%%%%%
\subsection{Fock space formalism}
Using complex coordinates $z_{\alpha}$,
we introduce the following operators:
\begin{equation}
 \hat{a} \equiv
  \frac{z}{\sqrt{\theta}}\;,\hspace{5mm}
 \hat{a}^{\dag}\equiv
  \frac{\bar{z}}{\sqrt{\theta}}
\;,\hspace{5mm}
 [\hat{a} , \hat{a}^{\dag}]
  = 1 \ , \ [\hat{a} , \hat{a}] =[\hat{a}^{\dagger} , \hat{a}^{\dagger}]=0\  
  \;.
\nonumber
\end{equation}
$\hat{a}^{\dag}$ is a creation operator and
$\hat{a}$ is an annihilation operator.
We define a
Hilbert space by
\begin{eqnarray}
 {\cal H}=\oplus {\bb C}\left|n \right> &,& \
 \left|n \right>=
  \frac{(\hat{a}^{\dag})^{n}}
  {\sqrt{n!}}\left|0 \right>,\nonumber\\
 \hat{a} \left|n \right>
  =\sqrt{n}\left|n-1 \right> &,&\ 
 \hat{a}^{\dag}\left|n \right>
  =\sqrt{n+1}\left|n+1 \right> ,
  \nonumber
\end{eqnarray}
where $\left|n \right>$ is a eigenvector of the number operator 
$\hat{n} \equiv\hat{a}^{\dag}\hat{a}$, i.e. 
$\hat{n} \left|k \right> = k \left|k \right>$.
An arbitrary operator has the following expression;
\begin{equation}
 \hat{\cal O}=\sum_{n,m} 
  {\cal O}_{m}^{n}
  \left|n \right>\left<m \right|\;  .
\nonumber
\end{equation}
Differentiation is given by
\begin{equation}
 \partial_{\mu}\hat{f}(\hat{x})=[\hat{\partial}_{\mu},\hat{f}(\hat{x})]
  =-i\theta^{-1} \epsilon_{\mu\nu}[\hat{x}^{\nu},\hat{f}(\hat{x})]\;.
\nonumber
\end{equation}
Here $\hat{\partial}_{\mu}= -i \theta^{-1} \epsilon_{\mu \nu}\hat{x}^{\nu}$ and 
$\epsilon_{\mu \nu}$ is the inverse of $\epsilon^{\mu \nu}$, i.e.
$\epsilon_{\mu\nu} \epsilon^{\nu\rho}=\delta^{\rho}_{\mu}$.
In terms of $\hat{a}$, $\hat{a}^{\dagger}$,
differentiation is expressed by
\begin{equation}
 \partial \hat{f}(z,\bar{z})=[\hat{\partial} , \hat{f}(\hat{x})]
  =-\frac{1}{\sqrt{\theta}} [ \hat{a}^{\dagger} , \hat{f}(z,\bar{z})]\; , \;
\bar{\partial} \hat{f}(\hat{x}) = [\hat{\bar{\partial}} , \hat{f}(z,\bar{z})]
=\frac{1}{\sqrt{\theta}} [ \hat{a} , \hat{f}(z,\bar{z})]\; .
\nonumber
\end{equation}

Integration is replaced by the trace operation,
\begin{equation}
\int d^2x \ f(x) = 2\pi \theta \mbox{Tr}_{\cal H} \hat{f}(\hat{x})
\nonumber
\end{equation}
in the operator formalism.
%%\end{slide}

The covariant derivative operator is defined by
\begin{equation}
\hat{\nabla}_{\mu}:=
\partial_{\mu} - i\hat{A}_{\mu}\; , \label{EQ:covariant-derivative2}
\end{equation}
where $\hat{A}$ is a gauge connection in the operator formalism.
For a Higgs field $\hat{\phi}$ in the operator formalism, the 
covariant derivative is given by
\begin{equation}
 \hat{\nabla}_{\mu}\hat{\phi}
  =[\hat{\partial}_{\mu},\hat{\phi}]- i \hat{A}_{\mu}\hat{\phi}
  =-\hat{\phi}\hat{\partial}_{\mu}
  +(\hat{\partial}_{\mu}- i \hat{A}_{\mu})\hat{\phi}\; ,
\end{equation}
where $\hat{\partial}_{\mu}=-i \theta^{-1} \epsilon_{\mu \nu}\hat{x}^{\nu}$.

%For convenience, let introduce operator $\hat{D}$ by
%\begin{equation}
% \hat{D}_{\mu}=\hat{\partial}_{\mu}- i \hat{A}_{\mu}\;,%\label{EQ:New-gauge}
%\nonumber
%\end{equation}
%this is an anti-hermitian operator and covariant under the 
%gauge transformation , 
%\begin{equation}
% \hat{D}_{\mu}\longrightarrow \hat{g}\hat{D}_{\mu}\hat{g}^{\dag}\;.
%\nonumber
%\end{equation}
The curvature is defined by
\begin{equation}
 \hat{F}_{\mu\nu}=i [\hat{\nabla}_{\mu},\hat{\nabla}_{\nu}]
%  =i\theta_{\mu\nu}+[\hat{D}_{\mu},\hat{D}_{\nu}]
\; ,
\label{kyokuritu}
\end{equation}
and the action functional of the gauge theory
in noncommutative ${\mathbb R}^{2}_{\theta}$ is given by
\begin{equation}
 S_{gauge}=- 2\pi \theta \frac{1}{2}
  \mbox{Tr}_{\cal H}\hat{F}_{\mu\nu}^2 \ .
%  \nonumber\\
%&=&
%-\frac{(2\pi)^d\sqrt{\det\theta^{\mu\nu}}}{4g^2}
%  \mbox{Tr}_{\cal H}\mbox{tr}_{U(N)}
%  \left([\hat{D}_{\mu},\hat{D}_{\nu}]+i\theta_{\mu\nu}\right)^2\; .
%\nonumber
\end{equation}

%%%%%%%%%%%%%%%%%%%%%%%%%%%%%%%%%%%%%%%%%%%%%%%%%%%%%%%%%%
%%%%%%%%%%%%%%%%%%%%%%%%%%%%%%%%%%%%%%%%%%%%%%%%%%%%%%%%%%

\subsection{An Explicit Solution}
%Consider the 2-dim case.
%Using complex coordinates $z$,
%creation and annihilation operators are given as one set by
%\begin{eqnarray}
% \hat{a} &\equiv&
%  \frac{z}{\sqrt{\theta}}\;,\hspace{5mm}
% \hat{a}^{\dag}\equiv
%  \frac{\bar{z}}{\sqrt{\theta}}
%\;,\hspace{5mm}
% [\hat{a} , \hat{a}^{\dag}]
%  = 1 \;.
%\nonumber
%\end{eqnarray}
%Hilbert space is expanded as
%\begin{eqnarray}
% {\cal H}=\oplus {\bb C}\left|n \right>\; &,& 
% \left|n \right>=
%  \frac{(\hat{a}^{\dag})^{n}}
%  {\sqrt{n!}}\left|0 \right>\;,\nonumber\\
% \hat{a} \left|n \right>
%  =\sqrt{n}\left|n-1 \right>\; &,&
% \hat{a}^{\dag}\left|n \right>
%  =\sqrt{n+1}\left|n+1 \right>\;.
%\nonumber
%\end{eqnarray}
%Number operator is 
%$\hat{n} \equiv\hat{a}^{\dag}\hat{a}$.
%Using the $\hat{a}$
%and $\hat{a}^{\dagger}$,
%the differentiation is expressed as
%\begin{equation}
% \partial \hat{f}(z,\bar{z})=[\hat{\partial} , \hat{f}(\hat{x})]
%  =-\frac{1}{\sqrt{\theta}} [ \hat{a}^{\dagger} , \hat{f}(z,\bar{z})]\; , \;
%\bar{\partial} \hat{f}(\hat{x}) = [\hat{\bar{\partial}} , \hat{f}(z,\bar{z})]
%=\frac{1}{\sqrt{\theta}} [ \hat{a} , \hat{f}(z,\bar{z})]\; .
%\nonumber
%\end{equation}
For a Higgs field $\hat{\phi}$ in the operator formalism, the 
covariant derivative has the complex expression : 
\begin{equation}
\hat{D} \hat{\phi} := [\hat{\partial} , \hat{\phi} ]-i \hat{A} \hat{\phi} 
 , \ \ \
\hat{\bar{D}} \hat{\phi} := [\hat{\bar{\partial}} , \hat{\phi} ]
                         - i \hat{\bar{A}} \hat{\phi} \ .
% \ , \\
%D * \bar{\phi}  = \partial \bar{\phi} + i \bar{\phi} * A  \ &,& \ \ \
%\bar{D} * \bar{\phi} = \bar{\partial}\bar{\phi} + i \bar{\phi} * \bar{A}
%\ \ .
\end{equation}
%where $\hat{\partial}_{\mu}= \theta_{\mu \nu}^{-1} \hat{x}^{\nu}$ and
%$\theta_{\mu \nu }^{-1}$ is the inverse of noncommutative
%parameter i.e. $\theta_{\mu \nu }^{-1} \theta_{\nu \rho} = \delta_{\mu \rho}$.

For $\hat{B}$ the magnetic field in operator formalism, 
we have
\begin{equation}
 \hat{B}=-i ([ \hat{\partial} , \hat{\bar{A}} ]
            - [ \hat{\bar{\partial}} , \hat{{A}} ]
            -[\hat{{A}} , \hat{\bar{A}} ])
%  =i\theta_{\mu\nu}+[\hat{D}_{\mu},\hat{D}_{\nu}] )
\; .
\label{hatB}
\end{equation}
In this formulation, the vortex equations are
\begin{eqnarray}
&&\hat{\bar{D}} \hat{\phi} =
[\hat{\bar{\partial}} , \hat{\phi}] -i \hat{\bar{A}}
\hat{\phi}
=
\frac{1}{\sqrt{\theta}} [\hat{a} , \hat{\phi}]
-i \hat{\bar{A}} \hat{\phi}=0
 \label{operatorBPS1} \\
&&\hat{B} + \hat{\phi} \hat{\bar{\phi}}-1=0
\label{operatorBPS2}
\end{eqnarray}
An explicit solution for (\ref{operatorBPS1}) and (\ref{operatorBPS2})
is given in \cite{Bak} by
\begin{equation}\label{solution}
\hat{\phi}= \sum_{n=0}^{\infty} | n+1 \rangle \langle n |
\; \; , \; \; 
\hat{A} = \frac{1}{i \sqrt{\theta}} 
\left( \hat{a} - \frac{\sqrt{\hat{n}}}{\sqrt{\hat{n}+1}} \hat{a} \right) .
\end{equation}
This solution has topological charge $\theta \mbox{Tr}_{\cal H} \hat{B} =1$.
In \cite{Bak}, explicit solutions are given for arbitrary integer valued 
topological charge $\theta \mbox{Tr}_{\cal H} \hat{B} =n$.
For simplicity, we discuss only (\ref{solution}).

We first translate the solution (\ref{solution})
into a $*$ product expression.
$|n \rangle \langle m|$ can be rewritten as
\begin{eqnarray}
|n \rangle \langle m| &=& : \frac{\hat{a}^{\dagger n}}{\sqrt{n!} }
e^{-\hat{a}^{\dagger} \hat{a}} \frac{\hat{a}^{m}}{\sqrt{m!}} : 
\nonumber \\
&=& \sum_{k=0}^{\infty} 
\frac{1}{\sqrt{n!m! \theta^{n+m}}}\left( \frac{-1}{\theta} \right)^k
\frac{1}{k!} \hat{z}^{k+m} \hat{\bar{z}}^{k+n} \; , 
\end{eqnarray}
where $: \sim : $ is normal ordering, which by
definition moves all $\hat{a}$'s to the right of
all the $\hat{a}^{\dagger}$'s.
%of the normal ordering is to put $\hat{a}$ 
%at $\hat{a}^{\dagger}$'s right hand.
{}From this fact, the $*$ product expression of $|n \rangle \langle m|$
is given by
\begin{equation}
\sum_{k=0}^{\infty} 
\frac{1}{\sqrt{n!m! \theta^{n+m}}}\left( \frac{-1}{\theta} \right)^k
\frac{1}{k!} {z}^{k+m} * \bar{z}^{k+n} \; .
\end{equation}
Therefore the Higgs field in the solution (\ref{solution}) is
\begin{eqnarray}
\phi &=& \sum_{n=0}^{\infty} \sum_{k=0}^{\infty} 
\frac{1}{(n!)\theta^n \sqrt{ (n+1) \theta}}\left( \frac{-1}{\theta} \right)^k
\frac{1}{k!} {z}^{k+n} * \bar{z}^{k+n+1}  \\
&=& e^{\frac{\theta}{2}\partial \bar{\partial} }
\varphi(z,\bar{z}) \ ,
\end{eqnarray}
where
\begin{equation}
\varphi(z,\bar{z}) := 
\bar{z} e^{-\frac{|z|^2}{\theta}}
\sum_{n=0}^{\infty} \frac{1}{n! \theta^n \sqrt{(n+1)\theta}} |z|^{2n} \; .
\label{varphi}
\end{equation}
By (\ref{varphi}), this type of solution has a
$1/\theta$ expansion, which differentiates solutions via Fock representation
from the solutions in section \ref{section vortex}.
Let us prove the following theorem.
%%%%%%%
\begin{thm}\label{conv_of_solution}
$ |\varphi | < 
 \int dx\  \frac{\sqrt{\theta}}{x^2} (1- e^{-\frac{x^2}{\theta}} )
$, where $|z|=x$.
\end{thm}

[Proof]
\begin{equation}
f(x):=
\sum_{n=0}^{\infty} \frac{1}{n! \theta^n \sqrt{(n+1)\theta}}x^{2n+1}
 , \ x \ge 0
\end{equation}
\begin{eqnarray}
\frac{d f(x)}{dx} -\frac{2x}{\theta} f(x)
&=& \frac{1}{\sqrt{\theta} }
+\frac{3-2\sqrt{2}}{\sqrt{2}}\frac{x^2}{\theta \sqrt{\theta}} \nonumber \\
&&+ \sum_{n=1}^{\infty} 
\frac{(2n+3)\sqrt{n+1}-(2n+2)\sqrt{n+2}}{(n+1)!\sqrt{(n+1)(n+2)}\theta^{n+1}\sqrt{\theta}} 
x^{2(n+1)} \nonumber \\
&<& \frac{1}{\sqrt{\theta}}\Big( 1 + \frac{x^2}{2! \theta}
+ \sum_{n=1}^{\infty} 
\frac{(4n^2-4n+9)x^{2(n+1)}}{(n+1)! \sqrt{n+2}\theta^{n+1} 8(n+1)^2(2n+1)}
\Big)
\nonumber \\
&<& \frac{1}{\sqrt{\theta}}\Big( 1 + \frac{x^2}{2! \theta}
+ \frac{\theta}{x^2} \sum_{n=1}^{\infty} 
\frac{x^{2(n+1)}}{(n+2)!\theta^{n+2}}
\Big)
\nonumber \\
&=& \frac{\sqrt{\theta}}{x^2} (e^{\frac{x^2}{\theta}} -1) .
\end{eqnarray}
Then,
\begin{equation}
| \varphi (x)|  = (e^{-\frac{x^2}{\theta}} f(x) ) 
 \le  \int dx\  \frac{\sqrt{\theta}}{x^2} (1- e^{-\frac{x^2}{\theta}} ) .
\end{equation}
\begin{flushright}
$\square $
\end{flushright}

This theorem shows that the existence of bounded solutions
with expansions in $1/{\theta}$.\\

\noindent
{\bf Acknowledgments}\\
Y.M. is supported by
21 century COE program: Integrative Mathematical Sciences, Progress in 
Mathematics Motivated by Natural and Social Phenomena,
and 
Partially supported by Grant-in-Aid for 
Scientific Research (\#18204006.), Ministry of Education , Science and 
Culture, Japan.

The authors appreciate for the helpful comments of the referee of Journal
of Geometry and Physics.

%%%%%%%%%%%%%%%%%%%%%%%%%%%%%%%%%%%%%%%%%%%%%%%%
%%%%%%%%%%%%%%%%%%%%%%%%%%%%%%%%%%%%%%%%%%%%%%%%

% \appendix

% \section{}
% \label{}

\end{document}